\documentclass[12pt]{article}

\usepackage{epsfig,amsmath,graphicx}

\newcommand{\muwa}{\mu_{\scriptscriptstyle\rm WA}}

\newcommand{\mupi}{\mu_\pi^2}
\newcommand{\mug}{\mu_G^2}
\newcommand{\rd}{\rho_D^3}
\newcommand{\rls}{\rho_{LS}^3}
\newcommand{\as}{\alpha_s}

\newcommand{\GeV}{\,\mbox{GeV}}

\def \be{\begin{equation}}
\def \ee{\end{equation}}
\newcommand{\bea}{\begin{eqnarray}}
\newcommand{\eea}{\end{eqnarray}}

\begin{document}
\begin{titlepage}

\begin{flushright}

\end{flushright}
\vskip 2cm

\centerline{\LARGE\bf\boldmath Normalizing  
inclusive rare $B$ decays}

\vskip 2cm

\begin{center}
{\bf 
  Paolo Gambino and  Paolo Giordano\\[3mm]
\it   Dip.\ Fisica Teorica, Univ.\ di Torino, \& INFN  Torino,
I-10125 Torino, Italy}

\end{center}

\vskip 2cm

\begin{abstract}
The inclusive semileptonic branching ratio is often employed to normalize 
other inclusive $B$ decays. Using recent determinations of the non-perturbative parameters of the Operator Product Expansion we compute the normalization factor for the branching ratio of $B\to X_s \gamma$
and find a few percent enhancement  with respect to previous determinations.

\end{abstract}

\end{titlepage}

\section{Introduction}
The partial width of the inclusive $B$ decays to light quarks depends on the fifth power of the $b$ quark mass. In order to reduce the  
uncertainty that follows from this strong sensitivity and to avoid the large radiative corrections that are sometimes related to the heavy quark mass, the Branching Ratio (BR) of rare decays is usually expressed in terms of the CKM favored semileptonic BR, BR$_{c\ell\nu}\equiv{\rm BR}[B\to X_c \ell \nu]$, a quantity that 
is presently known at the 1\% level \cite{HFAG,BF}. This choice
introduces a marked dependence on the charm quark mass in the calculation 
of rare decays,  but it is very convenient in $b\to s$ transitions, whose CKM factor $|V_{ts}^* V_{tb}|$ is essentially determined by 
$|V_{cb}| $ measured in inclusive semileptonic $B$ decays.
Moreover, in the case of $ B\to X_s \gamma$ the charm mass dependence 
due to the normalization to BR$_{c\ell\nu}$ is 
partially compensated by that of the perturbative QCD corrections.
In recent years the $B$ factories have performed increasingly detailed studies of semileptonic $B$ decays, providing us with improved determinations of the $b$ and $c$ quark masses and of the Operator Product Expansion (OPE)
parameters.  In parallel, both the
measurements and the theoretical calculations of inclusive rare decays have improved
significantly. In this Letter we reconsider the normalization of rare decays and 
try to assess its uncertainty, taking into account the latest developments. We will  concentrate on the radiative inclusive decay of the $B$ meson, $B\to X_s \gamma$: the Next-to-Next-to-Leading Order (NNLO) 
calculation of its BR in the Standard Model 
is quite advanced \cite{prl} and its experimental error will soon approach  5\%. Many of our considerations apply  to 
$ B\to X_s \ell^+\ell^-$ \cite{bobeth} and $B\to X_u \ell \nu$  as well.

As mentioned already, the unitarity of the CKM matrix implies 
$|V_{ts}^* V_{tb}|^2= [1 +\lambda^2 (2\bar \rho-1)+O(\lambda^4)]|V_{cb}|^2=(0.963\pm 0.003)\,|V_{cb}|^2 $ \cite{UTfit}, {\it i.e.}
the CKM prefactor of $ B\to X_s \gamma$
is essentially given by $|V_{cb}| $, whose best determination follows  
from BR$_{c\ell\nu}$ and has a 2\% error \cite{HFAG}. 
In recent analyses \cite{prl,gm,ms} the radiative BR is therefore expressed by
\be \label{one}
{\rm BR}_\gamma(E_0)\equiv
{\rm BR}[{B}\to X_s \gamma]_{ E_{\gamma} > E_0}= \frac{{\rm BR}_{c \ell \nu} }{C}\left(\frac{\Gamma[{B}\to X_s \gamma]_{ E_{\gamma} > E_0}}{
\left| {V_{cb}}/{V_{ub}} \right|^2   \Gamma[{B}\to X_u e \bar{\nu}]} \right),
\ee
where the phase space ratio $C$ is defined by
\be \label{phase2}
C = \left| \frac{V_{ub}}{V_{cb}} \right|^2 
\frac{\Gamma[{B} \to X_c e \bar{\nu}]}{\Gamma[{B} \to X_u e \bar{\nu}]}.
\ee
In Eq.(\ref{one}) the radiative width is normalized to the {\it charmless} semileptonic
decay in order to split  the charm mass dependence of the 
perturbative matrix elements of $b\to X_s \gamma$ --- an $O(\as)$  two-loop effect --- from that due to the normalization --- a tree-level effect, contained in $C$ \cite{gm}. This choice makes  the calculation more transparent.
One  has
\be \label{main}
\left(\frac{\Gamma[{B}\to X_s \gamma]_{ E_{\gamma} > E_0}}{
\left| {V_{cb}}/{V_{ub}} \right|^2   \Gamma[{B}\to X_u e \bar{\nu}]} \right)=
\left| \frac{ V^*_{ts} V_{tb}}{V_{cb}} \right|^2 
\frac{6 \,\alpha_{\rm em}}{\pi} 
\left[1 + \delta_{NP} \right]  P(E_0) ,
\ee
where $\alpha_{\rm em}$ is the fine structure constant. 
The perturbative term $P(E_0)$ is estimated  at NNLO in QCD \cite{prl,ms} and includes electroweak effects \cite{ewbsg}, while $\delta_{NP}$ contains the non-perturbative power corrections \cite{powerbsg}. Fortunately,
the $m_c$ dependence of $P(E_0)$, despite being a loop effect,  compensates about half of that of  $C$.

\section{Calculation of $C$}
Like all inclusive widths, the ratio $C$ can be calculated using the OPE and expressed as a double expansion in $\alpha_s$ and inverse powers of the $b$ quark mass, currently known 
through $O(\alpha_s^2)$ and $O(\Lambda_{\rm QCD}^3/m_b^3)$. $C$ depends sensitively on the $b$ and $c$ quark masses, as well as on the matrix elements of the dimension 5 and 6 operators. This is where the recent experimental studies 
of the inclusive moments of ${B}\to X_c e \bar{\nu}$ and $ B \to X_s\gamma$
enter in a crucial way. Indeed, the moments of various kinematic distributions 
provide information on the non-perturbative parameters of the OPE. Global fits
to the moments describe successfully a variety of moments and allow for a $40-50$MeV determination of $m_c$ and $m_b$, a $\sim 10-20\%$ determination of the 
$1/m_b^2$ and $1/m_b^3$ matrix elements, and a $\sim2\%$ determination of $|V_{cb}|$ \cite{BF,fit1s}. There are different ways to take into account the available information, relying on different assumptions and schemes. 
We work in the {\it kinetic} scheme \cite{kinetic}, where a `hard' cutoff $\mu$ separates perturbative and non-perturbative effects respecting heavy quark relations, and non-perturbative parameters are well-defined and perturbatively stable. 

Our starting point are 
the NNLO  expressions for the charmed and charmless total semileptonic widths 
\bea 
\label{crate}
\Gamma[\bar{B} \to X_c e \bar{\nu}] &=& \frac{G_F^2 \,m_b^5}{192\pi^3} |V_{cb}|^2
g(r) \left[ 1 +  \frac{\alpha_s}{\pi} p_c^{(1)}(r,\mu) 
              + \frac{\alpha_s^2}{\pi^2} p_c^{(2)}(r,\mu) 
\right. \nonumber\\ && \hspace{1cm} \left.            
  -\frac{\mupi}{2 m_b^2}               +         \left( \frac{1}{2} - \frac{2 (1-r)^4}{g(r)} \right) \frac{\mug-\frac{\rls+\rd}{m_b}}{m_b^2} 
              \right. \nonumber\\ && \hspace{1cm}\left.
           + \left(  8 \ln r   -\frac{10 r^4}{3}+\frac{32 r^3}{3}-8 r^2-\frac{32 r}{3}
            +\frac{34}{3} \right)
               \frac{\rd}{g(r) \,m_b^3}            
               \right],\\
\Gamma[\bar{B} \to X_u e \bar{\nu}] &=& \frac{G_F^2 \,m_b^5}{192\pi^3} |V_{ub}|^2
\left[ 1 +  \frac{\alpha_s}{\pi} p_u^{(1)}(\mu) 
              + \frac{\alpha_s^2}{\pi^2} p_u^{(2)}(r,\mu) 
              - \frac{\mupi}{2 m_b^2} - \frac{3\mug}{2m_b^2} 
 \right. \nonumber\\ && \hspace{1cm} \left.    
 +  \left(\frac{ 77}{6} + 8 \ln\frac{\muwa^2}{m_b^2}\right)
\frac{\rho _D^3}{ m_b^3} 
+\frac{3\rho _{LS}^3}{2 m_b^3} + \ \frac{32\pi^2}{m_b^3} 
B_{\scriptscriptstyle\rm WA}(\mu_{\scriptscriptstyle\rm WA})  
\right],
\label{urate} 
\eea
where $\alpha_s \equiv \alpha_s^{(n_f=5)}(m_b)$, $r =( m_c/m_b)^2$, $g(r) = 1 - 8 r + 8 r^3 - r^4 - 12 r^2 \ln r$, and all the masses and OPE parameters are defined
in the kinetic scheme at finite $m_b$ with  $\mu\sim 1$GeV.  
The non-perturbative corrections have been computed in \cite{power} and are expressed in terms of the parameters $\mu_\pi^2$, $\mu_G^2$, $\rd$, $\rls$.  
The matrix element of the Weak Annihilation (WA) operator 
$B_{\scriptscriptstyle\rm WA}\equiv
\langle B|O_{\scriptscriptstyle\rm WA}^u|B\rangle$ is poorly known. It is here 
renormalized in the $\overline{\rm MS}$ scheme at the scale $\muwa$, 
see \cite{WA,vub}.
We recall that  $B_{\scriptscriptstyle\rm WA}$ vanishes in the factorization approximation, and that WA is
phenomenologically important only to the extent factorization is
actually violated. There is however an $O(1)$ mixing between WA and 
Darwin operators, and 
at lowest order in perturbation theory one has 
$B_{\scriptscriptstyle\rm
WA}(\mu')=B_{\scriptscriptstyle\rm WA}(\mu)- \rd/2\pi^2 \, \ln \mu'/\mu$.
As factorization may hold only for a certain value  $\muwa=\mu_f$ 
for which  $B_{\scriptscriptstyle\rm WA}(\mu_f)=0$,  a
change of the scale $\mu_f$ provides a rough measure of the (minimal)
violation of factorization induced perturbatively. We neglect intrinsic charm contributions \cite{IC}. 
WA uncertainties make a precise prediction of $C$ problematic at present.
Fortunately,  they cancel out in Eq.(\ref{one}) since the radiative BR cannot depend on the  non-perturbative features of the charmless semileptonic decay. 

The perturbative corrections at $\mu=0$ (on-shell scheme) are given by
\bea
p_c^{(1)}(r,0) &=& -\frac{2 h(r)}{3 g(r)}, \ \ \ \ \ \ \ \ 
p_u^{(1)}(0) =\frac{25}{6} - \frac{2}{3} \pi^2, \label{pu1}\nonumber\\
p_c^{(2)}(r,0) &=& (-3.381 + 7.15 \sqrt{r} - 5.18\, r)\,
\beta_0^{(4)}   + (4.07-7.8 \sqrt{r}) 
,\label{pc2}\nonumber\\
p_u^{(2)}(r,0) &=& -3.22 \,\beta_0^{(4)} + 5.54 + 
(1.73 \ln \sqrt{r} -2.17)\sqrt{r} , \nonumber\label{pu2}
\eea
where $h(r)$ \cite{nir} is given in App.~C of \cite{gm}, and $\beta_0^{(n_f)} = 11 - \frac{2}{3} n_f$.
The $O(\as^2)$ perturbative corrections are known exactly in both cases
\cite{ritbergen, melnikov}. Their $m_c $ dependence is given here in terms of simple interpolation formulas, valid for $m_c/m_b$ between 0.2 and 0.3. 
While the $O(\as^2\beta_0)$ part of $p_c^{(2)}$ has been known for some time \cite{pc2blm}, 
the remaining (non-BLM) term is a very recent result \cite{melnikov}.
The $\mu$-dependence of $p_{u,c}^{(1,2)}(r,\mu)$ can be found exploiting the $\mu$-independence of the widths at each perturbative order and the known $\mu$-dependence of masses and OPE parameters 
\cite{Czarnecki:1997sz}
\bea\nonumber
&&m_q^{pole} \equiv m_q(0)= m_q(\mu) + \left[\overline{\Lambda}(\mu)\right]_{\rm pert}
+\frac{\left[\mu_{\pi}^2(\mu)\right]_{\rm pert}}{2m_q(\mu)}
\label{mrunn}\\
&&\mu_{\pi}^2(0) =  \mu_{\pi}^2 (\mu) - [\mu_{\pi}^2 (\mu)]_{\rm pert}\,,\  \ \ \ \ \
\rho_{D}^3(0)  =  \rho_{D}^3 (\mu) - [\rho_{\pi}^2 (\mu)]_{\rm pert}\, , \nonumber
\eea
where
\bea &
\left[\bar\Lambda(\mu)\right]_{\rm pert}= \frac{4}{3}C_F  
\frac{\alpha_s}{\pi} \mu 
\left(1+\frac{\as}{\pi} \left[\frac{\beta_0^{(3)}}{2}
\left(\ln{\frac{m_b}{2\mu}} + \frac{8}{3}\right)-
C_A\left(\frac{\pi^2}{6}-\frac{13}{12}\right)\right]\right),  
\nonumber\\
&
\left[\mu_{\pi}^2 (\mu)  \right]_{\rm pert} =\frac34 \mu 
\left[\bar\Lambda(\mu)\right]_{\rm pert}\!\!\! - \frac{C_F\as^2\beta_0^{(3)}}{\pi^2} 
\frac{\mu^2}4\,, \ \ \ \ \ \ \ \left[\rho_{D}^3 (\mu)\right]_{\rm pert} =\frac{\mu^2}2 \left[\bar\Lambda(\mu)\right]_{\rm pert}\!\!\! -\frac{C_F\as^2\beta_0^{(3)}}{\pi^2} \frac{2\mu^3}9
\nonumber
 \; , 
\eea
In the above formulas $C_F\!=\!\frac{4}{3}$,  $C_A\!=\!N_c\!=\!3$, and we 
have used $n_f=3$ in $\beta_0$, corresponding to three light massless quarks: indeed, the $m_c$ dependence of the charm loops on gluon lines is not known, and one can approximately decouple them. This is also consistent with the calculation of the moments in the kinetic scheme \cite{btoc}. A fully consistent $O(\as^2)$ implementation of the kinetic scheme would require the $O(\as)$ corrections to the Wilson coefficients of the higher dimensional operators in (\ref{crate},\ref{urate}), which is not yet available  except for the trivial $\mupi$ term. As for higher
order power corrections,  we recall that
$1/m_b^4$ corrections have been estimated to be tiny in the charmed decay rate \cite{1mb4}.
 
The numerical value of $C$ depends on those of the OPE parameters and of $\as$. We take as default values the results of the global fit \cite{BF}, namely
\bea\label{inputs}
m_b=4.597 \GeV, \ \ \ m_c=1.163\GeV, &&\ \ \  \mupi=0.436\GeV^2,\ \ \ \mug=0.267\GeV^2 \nonumber\\
\rd=0.213\GeV^3, &&\ \ \ \rls=-0.178\GeV^3,\nonumber
\eea
where all values are in the kinetic scheme with $\mu=1$GeV. We employ $\as(m_b)=0.219$ and set $\mu_{\scriptscriptstyle\rm WA}=m_b/2$, obtaining
\bea\label{Ckin1}
C&=& 0.546 -2.0 \,B_{\scriptscriptstyle\rm WA}\!\!\left(m_b/2\right)\\&=& 
0.625 - 0.028_{\as} - 0.022_{\as^2}
-0.004_{\mug}-0.025_{\rd} -0.001_{\rls} -2.0\, B_{\scriptscriptstyle\rm WA}\!\!\left(
m_b/2
\right)\nonumber
\eea
In the second line we have listed the individual contributions, that are accidentally all negative.
Using only BLM corrections and setting $n_f$ consistently equal to 3, one gets $C= 0.543-2.0 B_{\scriptscriptstyle\rm WA}(m_b/2)$, which shows that the $O(\as^2)$ corrections are dominated by $O(\as^2\beta_0)$ running coupling effects.
The $O(\as^2\beta_0)$ corrections can be absorbed in a rescaling of $\as$ in the NLO  contribution, using $\as(\mu_{BLM})$. It is clear however that in that case the BLM scale is very low, just above 1\GeV. 
Following \cite{vub} we use $B_{\scriptscriptstyle\rm WA}(m_b/2)\simeq0$ as central value and vary it between 0 and $+0.012\GeV^3$, since the positivity of the $q^2$  spectrum of $B\to X_u\ell \nu$ at $\muwa=m_b/2$ suggests a positive WA contribution. Taking into account the correlations among the input parameters \cite{BF}, and varying $\mu$ between 0.7 and 1.3\GeV\ to estimate the residual perturbative error, we find
\be\label{ckin2}
C=0.546\pm 0.017 (\rm par)\pm 0.016 (\rm pert) ^{+0.000}_{-0.024}(\rm WA)=0.546^{+0.023}_{- 0.033}.
\ee

This value can be compared with $C=0.580 +1.8\,B_{\scriptscriptstyle\rm WA}(m_b) \pm 0.016$, obtained in a global fit to semileptonic and  radiative moments \cite{fit1s} where the $b$ mass is in the $1S$ scheme \cite{1Smass}.  Using the  value of $\rd(\mu=0)$ obtained in \cite{fit1s}, we can rewrite  it as  $C=0.582 +
1.8\,B_{\scriptscriptstyle\rm WA}(m_b/2)$, that differs from Eq.(\ref{Ckin1}) by 6.5\%, exceeding the stated uncertainties.
We observe that the fit in \cite{fit1s} is based on older  data than that of \cite{BF}. The two fits  also differ in the perturbative scheme, in several assumptions, and in the estimate of theory errors, as detailed in Refs.~\cite{fit1s} and \cite{btoc,BF}, resp.  Ref.~\cite{fit1s}, for instance, does not extract the charm mass directly but eliminates it through the heavy meson mass  relation 
\be
m_b-m_c= \overline M_B-\overline M_D+\frac{\mupi}2 \left(\frac1{m_c}-\frac1{m_b}\right) +\frac{\rd-\bar\rho^3}4 \left(\frac1{m_c^2}-\frac1{m_b^2}\right) +O(1/m_Q^3),
\ee
 where all quark masses and OPE parameters have the same normalization point $\mu$. This relation  involves 
an expansion in $1/m_c$, rather than $1/m_b$, and some non-local operators (the term $\bar\rho^3$)
that do not enter the expressions for the semileptonic widths. In \cite{fit1s}
the normalization scale is set to zero and the fit is effectively equivalent to a direct fit to the pole mass difference $m_b-m_c$. 
The default procedure in \cite{fit1s} employs other meson mass relations
to fix $\mug$ and $\rls$, treats differently $\bar\Lambda$ terms, 
includes the moments of $B\to X_s \gamma$  accounting for neither distribution function effects nor an additional theory error, and estimates theory errors in a different way.

Despite their differences, the two fits give 
compatible values of $|V_{cb}|$.  But the ratio $C$, well approximated by  $ 1.2-2.2\, m_c/m_b$ in the physical region, 
is more sensitive to the exact value of the charm and bottom masses than $|V_{cb}|$, due to the correlation between $m_b$ and $m_c$ in the results of the fits. As illustrated in Fig.~1, the semileptonic rate and moments (in particular, those of the lepton energy distribution) stringently constrain 
a certain combination of $m_{c,b}$, given by $\approx m_b-0.65\,m_c$,
that roughly corresponds to constant 
values of the semileptonic total width for fixed $|V_{cb}|$.
In the fit of \cite{BF}  the latter gets  a 0.3\% relative parametric error, while $C\approx 1.2-2.2\, m_c/m_b$ has a 3.3\% error. The charmed inclusive width, 
roughly proportional to $m_b^5\, ( 1.2-2.2\, m_c/m_b)$, has instead a
 1.2\% relative uncertainty. 
As shown in Fig.~1, while the fitted value of $m_b^5\, ( 1.2-2.2\, m_c/m_b)$, and consequently of $|V_{cb}|$, is insensitive 
to small changes in the data and to the inclusion of radiative moments, the situation 
is quite different for $m_c$ and $C$. As indicated by the $1S$ fit to only Belle data 
performed by the Belle collaboration 
 \cite{belle}\footnote{The Belle data are also included in the global fit of \cite{BF}.},
the treatment of theoretical errors may similarly have a larger impact on the quark masses than on $|V_{cb}|$.
Different determinations of the $c$ and $b$ masses tend to prefer values close to the center of the plot: we show in Fig.~1 the 1$\sigma$ regions in the $(m_c,m_b)$ plane selected by the PDG  \cite{PDG} and by the $\sigma(e^+e^-\to{hadrons})$ sum rules according to \cite{steinhauser}, taking into account the non-negligible error introduced by the scheme translation (40 and 50 MeV  for $m_b$ and $m_c$, respectively, estimated using the residual scale dependence).

\begin{figure}[t]
\begin{center}
\includegraphics[width=11cm]{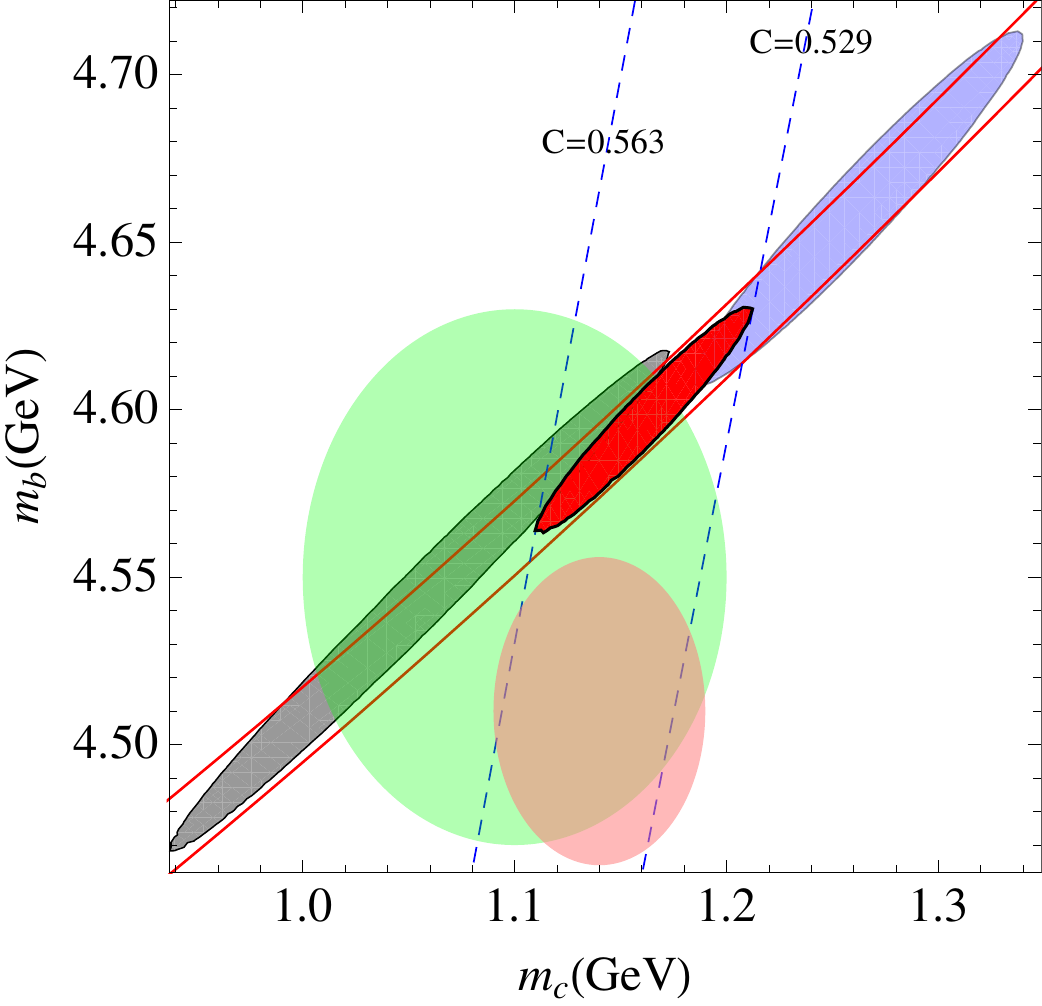}
\caption{\sf 1$\sigma$ contours in the $m_c,m_b$ plane 
in the kinetic scheme at $\mu=1\GeV$ from: $a)$ the global fit \cite{BF} (small red ellipse in the center); $b)$ same fit without the radiative moments (larger upper right ellipse);
$c)$ the Belle fit \cite{belle} (larger lower left ellipse); 
$d)$ the PDG  \cite{PDG} (large, light green central ellipse); $e)$
 the sum rules determinations  \cite{steinhauser} (smaller, light red, center bottom ellipse), after translating them in the kinetic scheme.
 The solid red and dashed blue lines correspond to constant values of the semileptonic width and of $C$, respectively. $C$ decreases moving to the right of the plot.}
\label{fig1}
\end{center}
\end{figure}

It is instructive to compare results based on the same set of data, like 
the recent Belle fits \cite{belle} that 
follow the methods of  \cite{btoc} and \cite{fit1s}. Using their kinetic scheme results,
in the same way that has led to Eq.(\ref{Ckin1}), we obtain $C=0.574$,  
although with a parametric error twice larger than above. 
From the central values of the 1S scheme fit of \cite{belle} we find $m_b^{pole}-m_c^{pole}= 3.393\GeV$ that can be employed to eliminate $m_c$ in Eq.(\ref{crate}), using the 1S scheme for $m_b$. This is analogous to what was done   in Appendix C of \cite{gm}, but includes $1/m_b^3$ effects that were neglected in that paper.
 Assuming $B_{\scriptscriptstyle WA}(m_b/2)=0$, the result is $C=0.563$.\footnote{Employing the same method with the results of \cite{fit1s}, we get $C=0.585$.}
In other words, using the same data set the results in the two schemes are closer and the kinetic scheme gives a higher value. The discrepancy between
Eq.(\ref{Ckin1}) and \cite{fit1s} is likely to be  due to different input data,
although at some level differences in the fit methods and in the treatment of  theory errors may also  play a role. 
The situation is likely to improve with better experimental data and better control of theory errors (an upgrade of the fitting methods is under way \cite{newcode}), but in the meantime the error attached to $C$ should be treated with caution.

\section{Impact on radiative decays}
We now move to the calculation of the radiative BR. The power correction $\delta_{NP}$ to the total inclusive radiative decay normalized to $B\to X_u \ell \nu$ \cite{powerbsg} --- see Eq.(\ref{main}) --- is
\be \label{ndel}
\delta_{NP}=
-\left(\frac{\mug}{27 m_c^2} +  \frac{\rd-\frac{13}{4}  \rls}{27 m_c^2 m_b}\right)\frac{\widetilde{C}^{(0)}(\mu_b)}{ C_7^{(0)\rm eff}(\mu_b)}  
- \left(\frac{44}3 +
8 \ln \frac{\mu^2_{\scriptscriptstyle\rm WA}}{m_b^2}\right)
\frac{\rd}{m_b^3} -\frac{32\pi^2}{m_b^3} B_{\scriptscriptstyle\rm WA}
(\mu_{\scriptscriptstyle\rm WA}),
\ee
where $\mu_b$ is the low-energy scale, that we set equal to 2.5\GeV\ like in \cite{ms}, $C_{7}^{(0)\rm eff}(\mu_b)\approx-0.37$ is the effective Wilson coefficients of the operator $Q_{7}$ calculated at leading logarithmic order, and  $\widetilde{C}^{(0)}(\mu_b)\approx1.20 $ is the appropriate combination of 
Wilson coefficients of $Q_{1,2}$ (see \cite{gm,ms}).
An additional $O(\as \mupi/m_b^2)$ contribution, dependent on the photon energy cut, has been calculated in \cite{neub}, but it is small for $E_0\leq 1.6\GeV$ and we neglect it. In the notation of  \cite{ms},  $\delta_{NP}=N(E_0)/P^{(0)}(\mu_b)$. 
Since the NNLO calculation of $P(E_0)$ \cite{prl,ms} has not been performed in the kinetic scheme,  all the parameters in Eq.(\ref{ndel}) must be understood at $\mu=0$, but we convert them to $\mu=1\GeV$ expanding to second order in $\as$.
We  find $\delta_{NP}=0.033-3.2 B_{\scriptscriptstyle\rm WA}
(m_b/2)$. Since $\delta_{NP}$  is correlated with $C$,  a  quantity that summarizes all power corrections and the normalization in Eq.\,(1) is
\bea\label{tot}
F&\equiv&\frac{1+\delta_{NP}}{C}=
1.600+ 0.075_{\as}+0.061_{\as^2} +0.047_{\mug}+0.057_{\rd}+
0.019_{\rls} \nonumber \\
&=&1.859\pm 0.054 (\rm par)\pm0.045(\rm pert)\, ,
\eea
which is independent of both  $B_{\scriptscriptstyle\rm WA}$ and $\mu_{\scriptscriptstyle\rm WA}$. 
As long as the perturbative corrections to Eq.(\ref{ndel}) are not available, the
numerical value of the charm mass to be employed is quite arbitrary. We have employed our default input value, but using $m_c\approx 1.5\GeV$, closer to the pole mass, would decrease $F$ by 1\%. We have accordingly increased the 
perturbative error in Eq.(\ref{tot}).
We list in the Appendix approximate formulae that allow for an easy determination of $C$ and $F$ with different inputs.
In the presence of new physics  $C_{7}^{(0)\rm eff}(\mu_b)=
C_{7, \rm SM}^{(0)\rm eff}(\mu_b)+C_{7,\rm new}^{(0)\rm eff}(\mu_b)$  and 
this is expected to be the only change in $F$, leading to 
\be
F=F_{\rm SM} - (0.06\pm 0.03) \, \varepsilon_{\rm new}
\ee
with $\varepsilon_{\rm new}= C_{7,\rm new}^{(0)\rm eff}(\mu_b)/C_{7}^{(0)\rm eff}(\mu_b)$ and $F_{\rm SM}$ given in Eq.(\ref{tot}). 

The BR with a  photon energy cut at the standard value $E_0=1.6\GeV$ can be computed from Eq.(\ref{one}) using the approximate relation
\be
\label{P}
P(1.6\GeV)\simeq  0.1247-0.0572 \left[m_c(m_c)-1.224\GeV\right]+0.0084\,(m_b^{1S}-4.68\GeV)
\ee
that can be extracted from \cite{ms}. Since this result is obtained with $m_b$ in the 1S  scheme and $m_c$ in the  $\overline{\rm MS}$ scheme, we  
perform an explicit change of scheme to the kinetic one, differentiating  
the NLO contribution to $P(E_0)$ wrt $m_{c,b}$.
Inserting Eqs.(\ref{tot},\ref{P})  into Eqs.(3) and (1)
and using BR$_{cl\nu}=0.1064$ from \cite{BF}, one gets
${\rm BR}_\gamma=3.28\times 10^{-4}$ which we choose as our central value.
Alternatively,  we can convert our kinetic scheme mass inputs 
to the 1S and $\overline{\rm MS}$ schemes and employ the results in Eq.~(\ref{P}), facing  however an extra theoretical error due to the conversion. 
As both masses enter $P(E_0)$  at $O(\as)$, 
one can use  one-loop formulas 
with $\as(2.5\GeV)\approx 0.27$, namely use $m_b^{1S}=4.69\GeV$  and $m_c(m_c)=1.23\GeV$ in Eq.~(\ref{P}).  This leads to 
${\rm BR}_\gamma=3.30\times 10^{-4}$.
If the conversion of  the kinetic masses to 
the 1S and $\overline{\rm MS}$ schemes is made via  two-loop
expressions,  ${\rm BR}_\gamma$ can  be as low as
$3.21\times 10^{-4}$, depending on the scale chosen for the $m_c$ conversion. 
 The spread of these values 
is an estimate of the perturbative higher orders consistent with that given in \cite{prl}. All renormalization scales involved in the calculation of $P(E_0)$ are kept 
to the default value of \cite{ms}.
The parametric error due to the OPE parameters and BR$_{cl\nu}$ 
given by the fit follows from the correlation matrix in \cite{BF} and amounts to only 1.8\%, to which we add  in quadrature the uncertainty from the CKM factor, $\alpha_s$, $m_t$, and the theory error in Eq.(\ref{tot}), obtaining 3.7\%. 
Summarizing, we have
\be
{\rm BR}_\gamma(1.6\GeV)=3.28\times 10^{-4}\left[1\pm 0.037 \pm 0.03\pm0.03\pm 0.05\right],
\ee
where the four errors are due to $i)$  the normalization and parametric uncertainties; $ii)$ the 
perturbative uncertainty in $P(1.6\GeV)$; $iii)$ the $m_c$ interpolation of \cite{ms}; $iv)$ unknown non-perturbative contributions beyond the OPE. Apart from $i)$, we have employed the same errors as in \cite{prl,ms} and 
our central value is about 4\%  higher than there.
 The main reason for the shift  is the new determination of $C$. Refs.\cite{prl,ms} employed  the most precise value of $C$ available at that time \cite{fit1s}.
Our total 3.7\% normalization and parametric error is a little larger than the corresponding uncertainty given in \cite{prl,ms}, but the total error is still dominated by the $\pm 5\%$ non-perturbative uncertainty.

The slow convergence of the perturbative series in Eqs.(\ref{Ckin1},\ref{tot}) is slightly disturbing. It is partly accidental, as both numerator and denominator have $\sim 2\%$ second order perturbative corrections, but with different sign. On the other hand,  
the kinetic mass definition is not particularly  appropriate for a quark with mass $m\sim \mu$. An alternative is to use a hybrid scheme where $m_b$ and the non-perturbative matrix elements are defined in the kinetic scheme, while the charm mass is defined in the $\overline{\rm MS}$ scheme. This yields a better apparent convergence in both $C$ and $\Gamma[B\to X_c e\bar\nu]$ if one employs $m_c(m_c)$, namely the  $\overline{\rm MS}$ scale of the charm mass is set equal to the mass itself.
As for the numerical value of $m_c(m_c)$, to simplify the comparison with Eqs.(\ref{Ckin1},\ref{tot}) we choose  $m_c(m_c)=1.267\GeV$ that is obtained from our input in (\ref{inputs}) through the two-loop perturbative relation with $\as(m_b)$. 
The results are 
\bea\label{Ckin2}
C&=& 0.542 -1.9\, B_{\scriptscriptstyle\rm WA}\!\!\left(m_b/2\right)\nonumber\\
&=& 0.574- 0.005_ {\as}  - 0.001_{\as^2} -0.004_{ \mug}-0.022_{\rd} -0.001_{\rls} - 1.9\,  B_{\scriptscriptstyle\rm WA}\!\!\left(m_b/2\right)\nonumber\\
F&=&1.886 
\\
&=&1.741+ 0.019_{\as}+0.003_{\as^2} +0.047_{\mug}+0.059_{\rd}+
0.018_{\rls} \nonumber 
\eea
which agree with  but converge better than Eqs.~(\ref{Ckin1},\ref{tot}). Since direct fits to $m_c(m_c)$ and kinetic scheme OPE parameters will soon become available \cite{newcode}, we list in the Appendix approximate formulae for this option too. The $\mu$-dependence  in Eqs.(\ref{Ckin2}) is less than 
$ 0.4\%$.
One can also calculate $C$ and $F$ using the $1S$ definition of the bottom mass \cite{1Smass} and the $\overline{\rm MS}$ scheme for $m_c$: using
again inputs that correspond to those employed in Eqs.(\ref{Ckin1},\ref{tot}), {\it i.e.}
$m_b^{1S}=4.74\GeV$  and $m_c(m_c)=1.267\GeV$, we find $C=0.545$ and $F=1.870$. Our results are quite stable for changes of scheme.

Finally, let us consider alternatives to the normalization of Eq.(1).
Since much of the $m_c$ sensitivity is related to the normalization to ${\rm BR}_{c l  \nu}$, one could hope to reduce it by avoiding 
${\rm BR}_{c \ell  \nu}$. This amounts to reexpressing the first ratio in Eq.(1)
 in terms of the semileptonic charmless width:
\be\label{norm2}
\frac{{\rm BR}_{c \ell  \nu}}{C}= \tau_B \left|\frac{V_{cb}}{V_{ub}}
\right|^2 \Gamma[ B\to X_u e \bar \nu]
\ee
where $\tau_B=1.585(7)$\,ps is the lifetime of  the $B^0/B^\pm$ admixture and $\Gamma[ B\to X_u e \bar \nu]$ is given in Eq.(\ref{urate}).  As $|V_{ub}|$ drops
out, the input parameters necessary to compute the rhs of Eq.(\ref{norm2}) are  $|V_{cb}|$, $m_b$, and the OPE parameters.
All these parameters, as well as $m_c$ and ${\rm BR}_{c \ell  \nu}$, can be extracted 
from a global fit as in \cite{BF}. Clearly in that case the
radiative BR will be the same and, because of the various correlations, will have the same error if computed using the rhs or the lhs of Eq.(\ref{norm2}). 
The rhs  is very sensitive to $m_b$ due to the $m_b^5$ factor in Eq.(\ref{urate}) but insensitive to $m_c$ (which however still enters  $P(E_0)$).
Moreover, the theoretical uncertainty in the 
extraction of $|V_{cb}|$ from ${\rm BR}_{c \ell  \nu}$ affects only the rhs of (\ref{norm2}).\footnote{See eq.~(7) in the second paper of Ref.~\cite{Czarnecki:1997sz} for an estimate.}
Neglecting all the correlations between the OPE parameters and BR$_{c\ell\nu}$, and using the rhs of Eq.(\ref{norm2}),
we get a $\sim 5\%$ parametric error instead of  $ 1.8\%$.  Even the very small errors on  $m_{c,b}$ found in \cite{steinhauser} would lead to a 3.3\%  uncertainty in this case.
This demonstrates the advantage of the normalization to the 
semileptonic BR and of a global fit to  the OPE parameters, as already stressed in \cite{ms}. In the future, independent and precise determinations of $m_b$ and $m_c$ could be used as additional constraints in the fit.  

\section{Summary}
We have calculated the normalization factor for radiative inclusive 
$B$  decays in the kinetic scheme and discussed its uncertainty and dependence on the input parameters, taking into account the correlations that arise from the fit 
to the moments of semileptonic and radiative distributions.
Using the latest global fit in the kinetic scheme \cite{BF} we obtain $C=0.546^{+0.023}_{- 0.033}$. We estimate that adopting our normalization in the 
first NNLO analysis of  \cite{prl,ms}  would lead to a 4\% higher BR$_\gamma$ for $E_\gamma>1.6\GeV$,  $(3.28\pm 0.25)\times 10^{-4}$.  
Since the  normalization factor is more 
sensitive to the value of the $c$ and $b$ quark masses than $|V_{cb}|$, its determination is presently more volatile, but  progress will come from improved measurements of the moments, a better understanding of the theoretical uncertainties involved in the fits, and complementary constraints on $m_{c,b}$. 
First steps to reduce the theory error in the normalization  
would be to employ the $\overline{\rm MS}$ charm mass and to compute higher order perturbative corrections, starting with the implementation of  \cite{melnikov}.

\subsection*{Acknowledgements}
We are grateful to M.~Misiak for many useful comments, suggestions and communications. We also thank H.~Fl\"acher, G.~Ossola, and P.~Urquijo for  useful communications.
This work is supported in part by  MIUR under contract 2004021808-009 and
by a European Community's Marie-Curie Research
Training Network under contract MRTN-CT-2006-035505 `Tools and Precision Calculations for Physics Discoveries at Colliders'.

\section*{Appendix}

The factors $C$ and $F$ defined in Eqs.(\ref{phase2},\ref{tot}) are well approximated by the following formulae, whose coefficients are given in Table 1.
\bea
C&= &g\left(r\right)\left[ c_1 + c_b \,\delta_b +c_c\, \delta_c+c_G \,\mug+c_D \,\rd+ c_{LS} \,\rls  +c_{\as} \, \delta_{\as}-\frac{32\pi^2}{m_b^3} B(m_b/2)\right]\nonumber \\
F&=& \left[g\left(r\right)\right]^{-1} \left[   c_1 + c_b \,\delta_b+c_c \,\delta_c+c_D\, \rd+\left(c_G\, \mug+ c_{LS}\, \rls\right)/m_c^2 +c_{\as} \, \delta_{\as}\right]
\nonumber 
\eea
\begin{table}[t]
\begin{center}
\begin{tabular}{|c|c|c|c|c|c|c|c|c|}
\hline
 & $m_c$ scheme & $c_1$ & $c_b$ & $c_c$ & $c_G$ &$c_D$ &$c_{LS}$& $c_{\as}$\\
 \hline
C & kin                     & 0.9185& 0.035& -0.001 &-0.021 & -0.186&  0.005& -0.53\\\hline
F &      kin                & 1.085 & -0.045& 0.007& 0.148& 0.169&  -0.091 & 0.60\\\hline
C & $\overline{\rm MS}$ & 1.001& 0.029 & -0.102 &-0.021 & -0.186&  0.005& 0.03\\\hline
F &   $\overline{\rm MS}$  & 1.001 & -0.035& 0.112& 0.148& 0.169&  -0.091& -0.01\\\hline
\end{tabular}
\caption{\sf Coefficients of the approximate formulae for $C$ and $F$ for different $m_c$ schemes.}
\end{center}
\label{default}
\end{table}%
Here all parameters are in the kinetic scheme with $\mu=1\GeV$, except for the 
charm mass that is either in the kinetic scheme or in the $\overline{\rm MS}$ scheme, $m_c(m_c)$. Moreover,  $\delta_b=m_b-4.6\GeV$,  $\delta_c=m_c-1.15\GeV$,  and $\delta_{\as}=\as-0.22$.
Notice that  in the range $0.22\le \sqrt{r}\le 0.29$, 
$g(r)\approx 1.1928-2.2443 \, m_c/m_b$ within 0.03\%.
The approximate formulae have a precision better than
0.3\% in the ranges $4.5< m_b<4.7\GeV$, $1<m_c<1.3\GeV$, $0.2<\as(m_b)<0.24$.

\end{document}